\documentclass[a4paper,11pt]{article}
\pdfoutput=1 

\usepackage{jheppub} 

\usepackage[T1]{fontenc} 

\usepackage{pdfsync}
\usepackage{mathptmx}
\usepackage[utf8]{inputenc}
\usepackage[T1]{fontenc}
\usepackage{slashed}
\usepackage{times}
\usepackage{amssymb,amsfonts,amsmath,amsthm}
\usepackage{dsfont,bbm}
\usepackage{dcolumn}
\usepackage{epsf}
\usepackage{graphicx}
\usepackage[caption=false]{subfig}
\usepackage{dsfont}
\usepackage{bm}
\usepackage{slashed}
\usepackage[active]{srcltx}
\usepackage[usenames]{color}
\newcommand{\up}{{\uparrow}}
\newcommand{\down}{{\downarrow}}

\title{\boldmath Alignment function as a new kind of transverse momentum dependent functions}

\author[a,b]{I.~V.~Anikin,}
\author[b]{L.~Szymanowski}

\affiliation[a]{Bogoliubov Laboratory of Theoretical Physics, JINR,\\141980 Dubna, Russia}
\affiliation[b]{National Centre for Nuclear Research (NCBJ),\\02-093 Warsaw, Poland}

\emailAdd{Igor.Anikin@ncbj.gov.pl}
\emailAdd{Lech.Szymanowski@ncbj.gov.pl}

\abstract{We argue the existence  of new  $k_\perp$-dependent functions 
which can be manifested in the Drell-Yan (SIDIS)-like processes.
The presented new functions resemble the well-known Boer-Mulders function
associated with the quark spin asymmetry, but in contrast they
are sensitive to the transverse motion of partons inside the hadron due to the 
collective alignment of quark spin vectors. }

\begin{document} 
\maketitle
\flushbottom

\section{Introduction}
\label{Intro}

Nowadays, the lepton production in nucleon-nucleon collisions in the Drell-Yan (DY) processes
attracts still much attentions 
not only of the experimental collaborations 
which investigate the composite (spin) structure of hadrons.
From the theoretical viewpoint, this kind of processes is interested because 
it opens the window for the studies of the transverse momentum dependent functions.
The most general form of the functions which depend on $k_\perp$ is given by
the hadron-hadron matrix element of quark-gluon operators
\footnote{The underlined Greek indices correspond to the open spinor indices;
$[0\,;z ]^{(\pm)}_A$ stands for the future- and past-pointed Wilson line (WL). 
Throughout the paper, we use the standard notations for the plus and minus light-cone directions.}
\begin{eqnarray}
\label{Phi-1}
\Phi^{(\pm)}_{\underline{\alpha} \underline{\beta}}(k) = 
\int (d^4 z) e^{+i(kz)} \langle P,S | \bar\psi_{\underline{\beta}}(0) \, 
[0\,;z ]^{(\pm)}_A \, \psi_{\underline{\alpha}}(z)  | P,S \rangle^H,
\end{eqnarray}
where $H$ indicates the Heisenberg representation (H-representation) used for the Fock states and operators.
Its general paramterization involving the hadron spin
axial-vector $S$ as one of Lorentz structures plays an important role for the study of different spin characteristics in 
the Drell-Yan (SIDIS)-like processes    
(see, for example, 
\cite{Boglione:1999pz, Bacchetta:2004jz, Goeke:2005hb, Collins:2005rq, Anselmino:2008sga, Bastami:2018xqd}).

The Lorentz parametrization of $\Phi^{[\Gamma]\,(\pm)}(k)$ with different Fierz projections is a extremely important stage 
for modern studies \cite{Goeke:2005hb} because, from the physical viewpoint, it defines the different sorts of parton distributions.
Among all parametrizing functions, the $k_\perp$-dependent  Boer-Mulders (BM) function introduced in \cite{Boer:1997nt}
and associated with the $\sigma^{+ \alpha} \gamma_5$-projection of $\Phi(k)$ in Eqn.~(\ref{Phi-1})
can be singled out as a function which 
describes the transverse spin 
asymmetry of quarks inside the unpolarized hadron. 
Indeed, for the BM-function $h^\perp_1(x, k^2_\perp)$ contribution,  
we have the following representation after the factorization procedure applied to the corresponding 
semi-inclusive deep-inelastic scattering (SIDIS)
\cite{Barone:2001sp}
\begin{eqnarray}
\label{BM-1}
&&\Phi^{[i\sigma^{+\alpha} \gamma_5]}(x, k^2_\perp) \sim 
{\cal P}^{q\down/N}(x, k_\perp) - {\cal P}^{q\up/N}(x, k_\perp)
\nonumber\\
&&
= \frac{|\vec{\bf k}_\perp|}{m_N} \sin(\phi_s-\phi_k) h^\perp_1(x, k^2_\perp),
\end{eqnarray}
where ${\cal P}^{q\down(\up) /N}(x, k_\perp)$ denotes the probability to find the transverse polarized quark inside the 
unpolarized hadron,
$\phi_s-\phi_k$ defines the angle between $\vec{\bf k}_\perp$ and $\vec{\bf s}_\perp$
being denoted as the quark transverse momentum and quark transverse spin vector (in two dimensional Euclidian space) respectively.
The expression $|\vec{\bf k}_\perp|\sin(\phi_s-\phi_k)$  in Eqn.~(\ref{BM-1}) stems from the vector product $\vec{\bf k}_\perp \wedge \vec{\bf s}_\perp$ 
provided $|\vec{\bf s}_\perp| =1$.

The aim of the paper is to study a new manifestation of $k_\perp$-dependent parameterizing functions 
of the matrix element (\ref{BM-1})
which can be associated with the inner transverse 
quark motion generated by the collective spin alignment rather than the quark spin asymmetry.
In other words, the functions we consider are related to the probability ${\cal P}^{q\up_x/N}(x, k_\perp)$ defined by
the operator
$[ \bar\psi^{(\up_x)} \gamma^+\gamma_1 \gamma_5 \psi^{(\up_x)}]$ involving 
the projections $\psi^{(\up\down_i)}=1/2 (1 \pm \gamma_i \gamma_5) \psi$.
Here, $(i=1)\Leftrightarrow (x)$, $(i=2)\Leftrightarrow (y)$ and $x$, $y$ are the polarization axes).

\section{New transverse momentum dependent functions}
\label{NF}

In this section, we introduce new transverse momentum dependent functions which are associated 
with the spin structure of quark content of hadrons.

To begin with, 
we rewrite Eqn.~(\ref{Phi-1}) in the interaction representation (see Appendix~\ref{App-Amp} for details) focusing {\it e.g.}
on the vector projection
\footnote{The symbol of time-ordering and the Wilson line are omitted}
\begin{eqnarray}
\label{Phi-1-2}
\Phi^{[\gamma^\mu]}(k) = 
\int (d^4 z) e^{+i(kz)} \langle P,S | \bar\psi(0) \, 
\gamma^\mu\, \psi(z) \, \mathbb{S}[\psi,\bar\psi, A] \,| P,S \rangle, 
\end{eqnarray} 
where $P$ and $S$ are the momentum vector and spin axial-vector of hadron, respectively, and 
\begin{eqnarray}
\label{S-m}
\mathbb{S}[\psi,\bar\psi, A]=T\text{exp}\Big\{ 
i \int (d^4 \xi) \big[ {\cal L}_{QCD}(\xi) + {\cal L}_{QED}(\xi)\big] \Big\}.
\end{eqnarray}
For brevity, all corresponding normalization and dimensionful pre-factors have been absorbed in the 
definitions of integration measures or in the definitions of parametrizing functions.
Notice that the function $\Phi^{[\gamma^\mu]}(k)$ is a very typical one which appears to describe the 
different processes in (semi)inclusive and (semi)exclusive channels.

\subsection{Lorentz parametrization of $\Phi^{[\gamma^\mu]}(k)$: the ``standard'' way}
\label{LP-st}

\begin{figure}[tbp]
\centering 
\includegraphics[width=.45\textwidth]{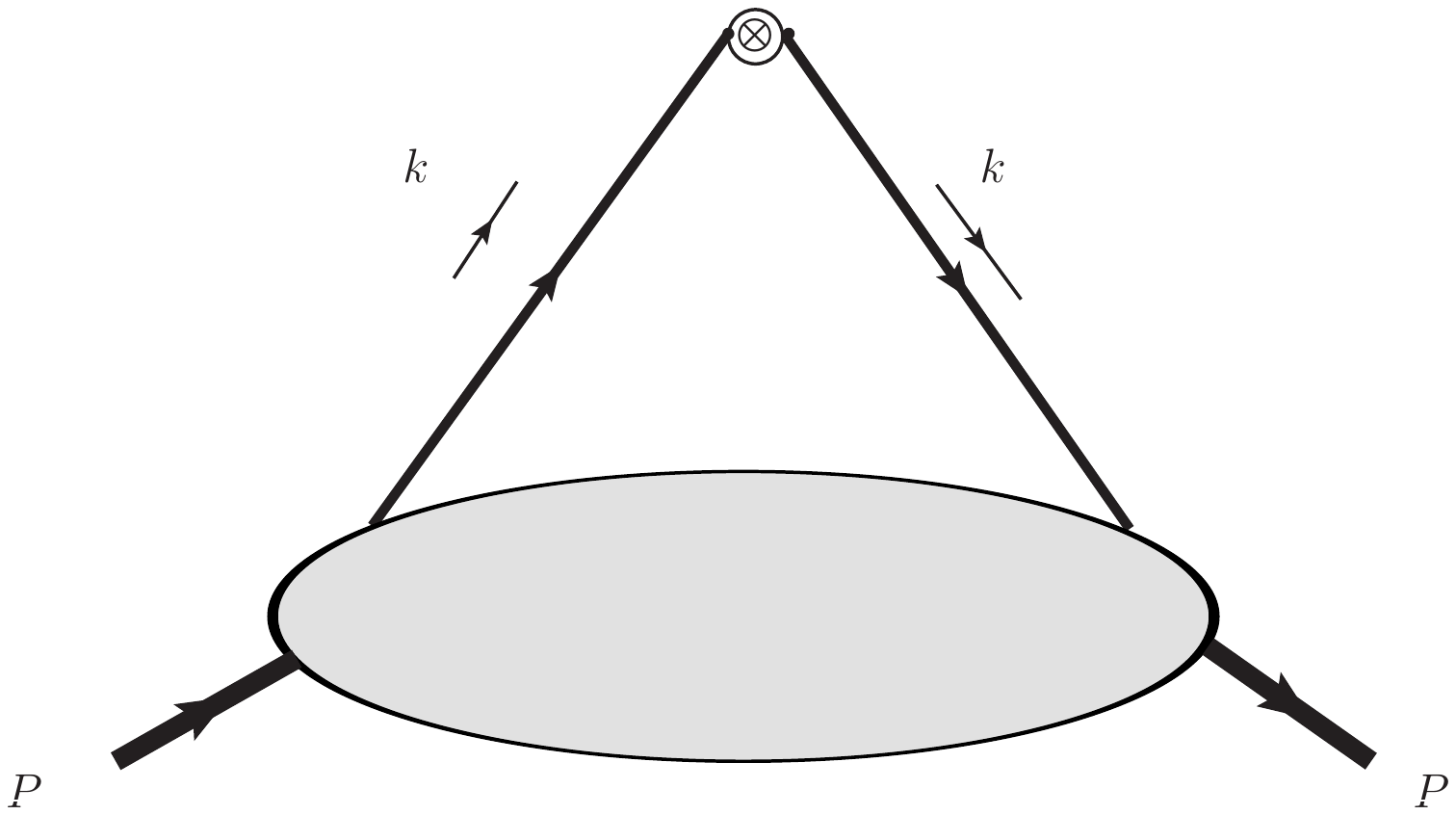},
\includegraphics[width=.45\textwidth]{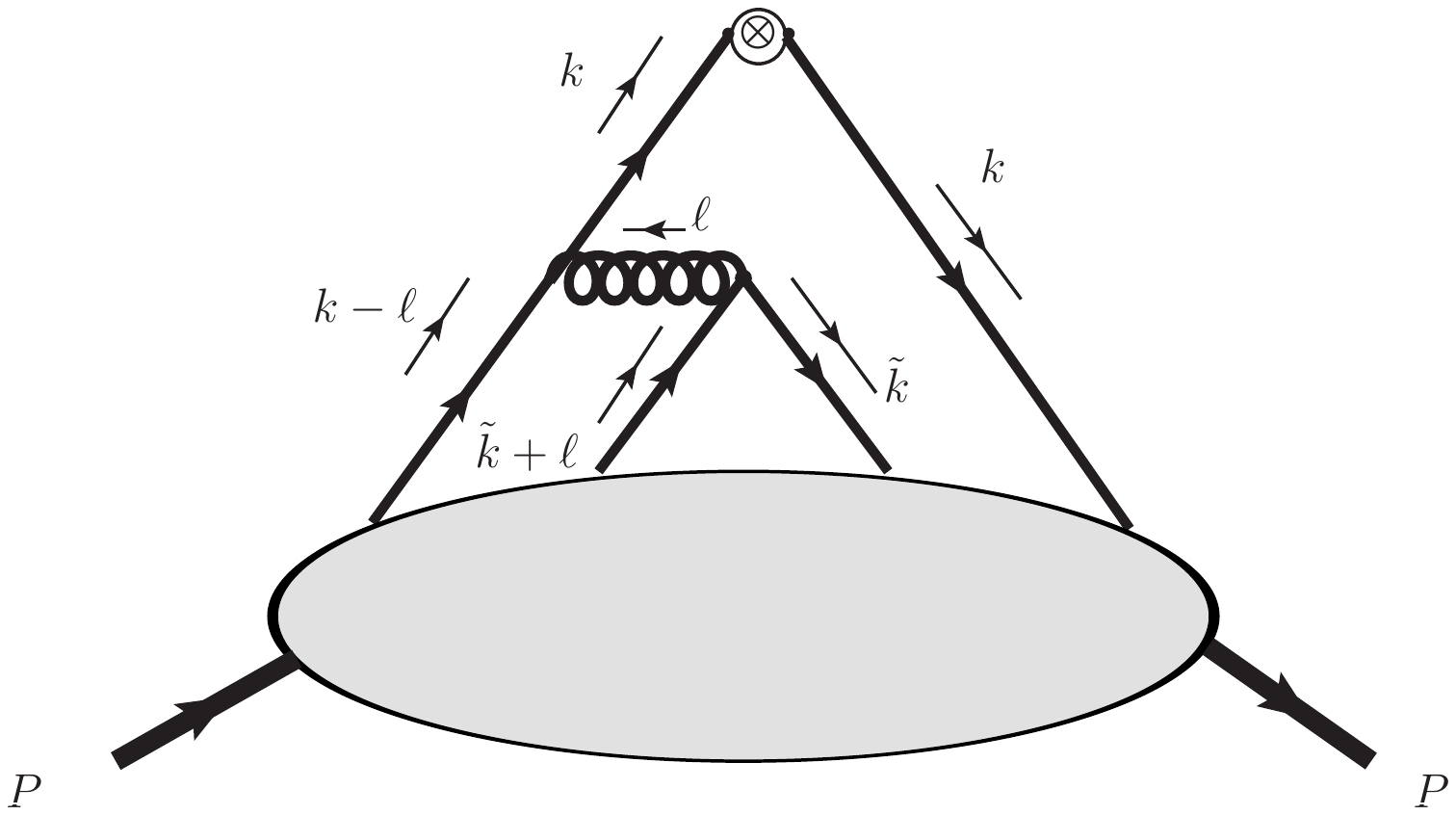}
\vspace{-5cm}
\caption{\label{Fig-S-1} The vector correlator with $\mathbb{S}=\mathbb{I}$ (the left panel); 
The vector correlator with $\mathbb{S}^{(2)}$ (the right panel).}
\end{figure}

The Lorentz decomposition (or parameterization) of any relevant correlators is usually based on the 
following standard scheme:  
\begin{itemize}
\item the parton operators in correlators are supposed to be considered as free operators, 
{\it i.e.} $\mathbb{S}[\psi,\bar\psi, A] = \mathbb{I}$, see Fig.~\ref{Fig-S-1}, the left panel; 
\item one forms the orthogonal system of Lorentz tensors;
\item with the help of the basis Lorentz tensors, one parametrizes the given Fierz projection of correlators 
based on the principle of Lorentz covariance;
\item including $\mathbb{S}$-matrix in the correlator, 
one studies the corresponding evolution of parametrizing functions. 
\end{itemize}

Within this standard scheme, the most general parameterizations
of relevant correlators can be found, for example, in  \cite{Barone:2001sp, Goeke:2005hb}.
For illustration, let us dwell on the following decomposition which includes the transverse momentum dependence of 
quarks:
\begin{eqnarray}
\label{Phi-1-3}
\Phi^{[\gamma^+]}(k)\Big|_{\mathbb{S}[\psi,\bar\psi, A]=\mathbb{I}} = P^+ f_{1} \left(x; \, k^2_\perp, (k_\perp P_\perp)\right). 
\end{eqnarray} 
It is important to emphasize that the quark combination 
$\langle P, S | \bar\psi \gamma^+ \psi | P, S\rangle$ of Eqn.~(\ref{Phi-1-3})
singles out the unpolarized quarks only.
In addition to $P^+$-structure, we may construct (without thinking on the orthogonality) the Lorentz vector 
$i\, \varepsilon^{+- k_\perp S_\perp}$  
for parametrization of $\Phi^{[\gamma^+]}(k)$ in Eqn.~(\ref{Phi-1-3}) even for the unpolarized quark combination
\footnote{$S_\perp$ is the covariant spin axial-vector of transversely polarized hadron}. 
Indeed, in this case, $S_\perp$ plays a role of the exterior axial-vector which is not dictated by the quark combination.
It leads to the term with the well-known function $f^\perp_T(x, k_\perp)$ in parametrization of $\Phi^{[\gamma^+]}(k)$,
{\it i.e.} we go over to the following parametrization \cite{Goeke:2005hb} 
\footnote{Modulo the prefactors which are irrelevant for our study.}
\begin{eqnarray}
\label{Phi-1-4}
\Phi^{[\gamma^+]}(k)\Big|_{\mathbb{S}[\psi,\bar\psi, A]=\mathbb{I}} = 
P^+ f_{1} \left(x; \, k^2_\perp, (k_\perp P_\perp)\right) + i\, \varepsilon^{+ - k_\perp S_\perp}
f^\perp_T\left(x; \, k^2_\perp, (k_\perp P_\perp)\right).
\end{eqnarray} 
However, the Lorentz vector $i\, \varepsilon^{+- k_\perp s_\perp}$ with the quark spin axial-vector
\footnote{Based on the Lorentz covariance, the combination $i\, \varepsilon^{+- k_\perp s_\perp}$ as 
the Lorentz vector is not formally excluded from the parametrization of the vector correlator.}
is  absent for the unpolarized quarks in correlator of Eqn.~(\ref{Phi-1-4}) 
due to the trace given by 
\begin{eqnarray}
\label{Tr-unp-q}
&&- \langle P, S | \bar\psi^{(s)} \gamma^+ \psi^{(s)} | P, S\rangle = 
\langle P, S | \text{tr} \big[ \gamma^+ \psi^{(s)} \bar\psi^{(s)} \big] | P, S\rangle=
\nonumber\\
&&\langle P, S | \text{tr} \big[ \gamma^+ (\hat k + m_q) (1+ \gamma_5 \lambda_s + 
\gamma_5 \hat s_\perp) \big] | P, S\rangle.
\end{eqnarray}
All the material presented in this subsection is not surprised and well-known in the modern literature \cite{Barone:2001sp, Goeke:2005hb}.
However, the principle lack of this method is that the Lorentz parametrization (or Lorentz decomposition) 
is implemented forgetting, in a sense, the presence of interaction in the correlator.

%
%
%

\subsection{Lorentz parameterization of $\Phi^{[\gamma^\mu]}(k)$: the ``nonstandard'' way}
\label{LP-nonst}

In this subsection we propose an alternative way to parameterize the corresponding 
correlators which opens a window for the new kind of functions.
At the very beginning, we take into account that the correlator involves the corresponding interaction 
(see, for example, Fig.~\ref{Fig-S-1}, the right panel).
Namely,  we propose to follow the scheme defined by
\begin{itemize}
\item the parton operators in correlators are assumed to be considered as interacting operators, 
{\it i.e.} $\mathbb{S}[\psi,\bar\psi, A]$ is presented in the correlator of Eqn.~(\ref{Phi-1-2}); 
\item one forms the orthogonal system of Lorentz tensors;
\item with the help of the basis Lorentz tensors, one parametrizes the given Fierz projection of correlators 
based on the principle of Lorentz covariance;
\item taking into account that $\mathbb{S}$-matrix is already presented in the correlator, 
one studies the corresponding evolution of parametrizing functions. 
\end{itemize}  

In other words, the substantial difference between the standard and nonstandard ways is the following:
in the standard way, we first consider free operators to parametrize the given correlator
and $\mathbb{S}$-matrix has been included in the consideration to derive the evolution
of the already-introduced parametrizing functions;
while, in the nonstandard way, we parametrize the given correlator where $\mathbb{S}$-matrix has been 
presented in Eqn.~(\ref{Phi-1-2}) from the very beginning. 

We believe that the nonstandard way is more adequate because, first, $\mathbb{S}$-matrix is
presented in the correlator even before factorization and, second, as shown below,
one can discover the new structure functions which are invisible in the standard way. 

Using the nonstandard way, we cannot exclude the following new structure:
\begin{eqnarray}
\label{Phi-plus-0}
&&\Phi^{[\gamma^+]}(k) = i \epsilon^{+ - P_\perp s_\perp} \tilde f_1^{(1)} (x; \,k^2_\perp) + ....
\end{eqnarray} 
where $s_\perp$ stands for the quark spin axial-vector and $\mathbb{S}$-matrix presents in the correlator.
The details of derivation are collected in the next subsection.   
 
 \subsection{Derivation of the new $k_\perp$-dependent functions}
 \label{Der-NF}

 We are gong over to the technical details of consequences of the nonstandard scheme.
 Let us focus on
the well-know $k_\perp$-dep\-endent function $f_1$ which parameterizes
the unpolarized hadron matrix element of quark operator involving $\gamma^+$ as 
\begin{eqnarray}
\label{Me-1}
&&
\Phi^{[\gamma^+]}(k)=
P^+ f_{1} \left(x; \, k^2_\perp, (k_\perp P_\perp)\right) =
\\
&&
P^+ (k_\perp P_\perp) f_1^{(1)} (x; \,k^2_\perp) + 
\Big\{ \text{terms of} \,\,(k_\perp P_\perp)^n  \,| \, n=0, n\ge 2  \Big\},
\nonumber
\end{eqnarray}
where $k=(xP^+, k^-, \vec{\bf k}_\perp)$ and $f_{1} \left(x; \, k^2_\perp, (k_\perp P_\perp)\right)$ 
has been decomposed into the powers of $(k_\perp P_\perp)$. 
Keeping the term of decomposition with $n=1$ represents the minimal necessary reqirement for 
the manifestation of new functions. 

Notice that $P_\perp$ is substantially non-zero within the Collins-Soper (CS) frame of Drell-Yan-like (DY) processes.
The matrix elements in Eqns.~(\ref{BM-1}) and (\ref{Me-1}) are usually treated within the 
H-representation (see Appendix~\ref{App-Amp}) whereas,  
for our goals, we mainly adhere the interaction representation (I-representation).
It is worth to notice that the Lorentz structures which parameterize the corresponding correlators
are related to the spinor lines appearing in consideration at the given order of the coupling constant,
see below Eqns.~(\ref{Demo-1})-(\ref{Demo-2}).

After performing the Fourier transforms of the quark operators in (\ref{Me-1}), we obtain that 
\begin{eqnarray}
\label{Me-1-2}
&&\int (d^4 k) e^{-i(kz)} \Phi^{[\gamma^+]}(k)=
\nonumber\\
&&
\int d\mu^{[\gamma^+]}_{\lambda\, \lambda^\prime} (k_1, k_2)
\langle P,S | b^+_{\lambda}(k_1) b^-_{\lambda^\prime}(k_2) | P,S \rangle^H,
\end{eqnarray} 
where 
\begin{eqnarray}
\label{mi-1}
d\mu^{[\gamma^+]}_{\lambda\, \lambda^\prime} (k_1, k_2) = 
(d^4 k_1) (d^4 k_2)e^{-i (k_2 z)}
\big[  \bar u^{(\lambda)} (k_1) \gamma^+ u^{(\lambda^\prime)}(k_2)\big] 
\end{eqnarray}
and
\begin{eqnarray}
\label{Me-1-3}
&& 
\langle P,S | b^+_{\lambda}(k_1) b^-_{\lambda^\prime}(k_2) | P,S \rangle^H=
\langle P,S | b^+_{\lambda}(k_1) b^-_{\lambda^\prime}(k_2) \, \mathbb{S}[\psi,\bar\psi, A] \, | P,S \rangle \equiv
\nonumber\\
&&\delta^{(4)} (k_1-k_2) {\cal M} _{\lambda\, \lambda^\prime}(k_1^2, k_2^2, (k_1 k_2) | P),
\end{eqnarray} 
where $b^{\pm}$ are the quark creation and annihilation operators.

In what follows we omit the spin state indices $\lambda$ and $\lambda^\prime$ and 
we do not pay an attention on the difference between contra- and covariant vectors
unless it may lead to a confusion.

Taking into account the delta function of (\ref{Me-1-3}), we can see that the 
spinor line formed by $\big[  \bar u (k) \gamma^+ u(k)\big]$ results in the appearance of  $k^+\sim P^+$
in the most trivial case without spin polarizations and interactions.

We now consider 
the second order of strong interactions in the correlator within the interaction representation, see Fig.~\ref{Fig-S-1}, the right panel. 
We have the following expression 
\begin{eqnarray}
\label{Me-2}
&&
\langle {\cal O}^{[\gamma^+]}\rangle^{(2)} \equiv
\langle P,S | T \bar\psi(0) \gamma^+ \psi(z) \, \mathbb{S}_{QCD}^{(2)}[\psi,\bar\psi, A] \,| P,S \rangle =
\nonumber\\
&&
\int (d^4 k) e^{-i (kz)} \Delta(k^2) \int (d^4 \ell) \Delta(\ell^2) \int (d^4 \tilde k) 
\nonumber\\
&&
\times {\cal M}\left( k^2, \ell^2, \tilde k^2, ... \right) 
\nonumber\\
&&\times
\big[ \bar u (k) \gamma^+ \hat k \gamma^\perp_\alpha u(k-\ell)\big]
\big[ \bar u (\tilde k) \gamma^\perp_\alpha u(\tilde k + \ell)\big],
\end{eqnarray}
where $\mathbb{S}^{(2)}_{QCD}$ denotes the $\mathbb{S}$-matrix operator at the order of $g^2$.
In Eqn.~(\ref{Me-2})
the quark and gluon propagators read
\begin{eqnarray}
\label{q-g-prop}
&&S(k)=\hat k \Delta(k^2), \quad  D^\perp_{\mu\nu}(\ell)= g^\perp_{\mu\nu} \Delta(\ell^2),
\nonumber\\
&&
\Delta(k^2) = \frac{1}{k^2 + i\epsilon}, \quad \hat k = (k\gamma)
\end{eqnarray}
and the amplitude ${\cal M}$ is given by  
\begin{eqnarray}
\label{M-me}
&&{\cal M} \left( k_i^2,  (k_i k_j), ...\right) \delta (k_1+k_3 - k_2-k_4) =
\nonumber\\
&&
\langle P,S | b^+(k_1) b^-(k_2) b^+(k_3) b^-(k_4) | P,S \rangle.
\end{eqnarray}
We single out the region where $| \ell | \ll \{|k|, |\tilde k| \}$ and as a result we obtain that 
\begin{eqnarray}
\label{Me-3}
&&
\langle {\cal O}^{[\gamma^+]}\rangle^{(2)} 
\sim \int (d^4 k) e^{-i (kz)} \Delta(k^2) 
\big[ \bar u (k) \gamma^+ \hat k \gamma^\perp_\alpha u(k)\big]
\\
&&
\times 
\int (d^4 \tilde k) \big[ \bar u (\tilde k) \gamma^\perp_\alpha u(\tilde k )\big]
\int (d^4 \ell) \Delta(\ell^2)  
{\cal M}\left( k^2, \ell^2, \tilde k^2, ... \right).
\nonumber
\end{eqnarray}
The next stage is to transform the spinor lines of this expression. For the first spinor line, we write  
\begin{eqnarray}
\label{Sp-line-1} 
&&\big[ \bar u (k) \gamma^+ \hat k \gamma_\perp^\alpha u(k)\big] = 
{\cal S}^{+ k \alpha \beta} \big[ \bar u (k) \gamma^\beta u(k)\big] + (\text{axial})
\nonumber\\
&&\Longrightarrow  k_\perp^\alpha \big[ \bar u (k) \gamma^+ u(k)\big]
+ (\text{other terms}),
\end{eqnarray}
where the following notation has been used
\begin{eqnarray}
{\cal S}^{\mu_1 \mu_2 \mu_3 \mu_4} =\frac{1}{4} \text{tr} \big[
\gamma^{\mu_1}\gamma^{\mu_2}\gamma^{\mu_3}\gamma^{\mu_4} \big].
\end{eqnarray}
The second spinor line can be considered with the help of the covariant (invariant) integration given by 
\begin{eqnarray}
\label{Sp-line-2}
&&k_\perp^\alpha
\int (d^4 \tilde k) \,\big[ \bar u (\tilde k) \gamma^\alpha u(\tilde k )\big]\,
{\cal M}\left( \tilde k^2, (\tilde k P), ... \right)=
\nonumber\\
&&
k_\perp^\alpha
\int (d^4 \tilde k) \,\tilde k^\alpha\,
{\cal M}\left( \tilde k^2, (\tilde k P), ... \right)=
\nonumber\\
&&
(k_\perp P_\perp) \int (d^4 \tilde k) \,\frac{(\tilde k_\perp P_\perp)}{P^2_\perp}\,
{\cal M}\left( \tilde k^2, (\tilde k P), ... \right).
\end{eqnarray}
Using (\ref{Sp-line-1}) and (\ref{Sp-line-2}), one can see that the form of  the {\it r.h.s.} of (\ref{Me-3})
coincides with the parametrization of (\ref{Me-1}) at $g^2$-order. 
Indeed, we have the following
\begin{eqnarray}
\label{Me-4}
&&
P^+ (k_\perp P_\perp) f_1^{(1)} (x; \,k^2_\perp) \sim
\nonumber\\
&& 
\big[ \bar u (k) \gamma^+ u(k)\big]
\int (d^4 \tilde k) \big[ \bar u (\tilde k) \hat k_\perp u(\tilde k )\big]
\nonumber\\
&&
\times
\int (d^4 \ell) \Delta(\ell^2)  \Delta(k^2)
{\cal M}\left( k^2, \ell^2, \tilde k^2, ... \right).
\end{eqnarray}
In other words, the parametrization of (\ref{Me-1}) with the Lorentz combination $P^+ (k_\perp P_\perp)$
is related to the two spinor lines
\begin{eqnarray}
\label{Lor-vec-1}
\big[ \bar u (k) \gamma^+ u(k)\big] \Rightarrow k^+ \sim P^+, 
\quad \big[ \bar u (\tilde k) \hat k_\perp u(\tilde k )\big] \Rightarrow (k_\perp P_\perp)
\end{eqnarray}
at $g^2$-order.

In the region of $|\tilde k| \sim |k|$, 
two spinor lines of (\ref{Me-4}) can be transformed into the other spinor lines 
with the help of Fierz transformations.
Using its general form (see, \cite{Itzykson:1980rh})
\begin{eqnarray}
\label{Gen-F-tr}
&&\big[ \bar u^{(a)} O_1  u^{(b)} \big]
\big[ \bar u^{(c)} O_2  u^{(d)} \big]=
\nonumber\\
&&
\frac{1}{4} \sum_{A, R_1, R_2}
\Big\{ \frac{1}{4} \text{tr} \big[
\Gamma_A O_1 \Gamma_{R_1} 
\big] 
\Big\}
\Big\{ \frac{1}{4} \text{tr} \big[
\Gamma^A O_2 \Gamma_{R_2} 
\big] 
\Big\}
\nonumber\\
&&
\times 
\big[ \bar u^{(c)} \Gamma^{R_1}  u^{(b)} \big]
\big[ \bar u^{(a)} \Gamma^{R_2}  u^{(d)} \big]
\end{eqnarray}
with 
$O_1= \gamma^+ \gamma^\perp_j \gamma_5$, 
$O_2={\bf 1}$,
$\Gamma^A= \gamma_i^\perp$, $\Gamma^{R_1}=\gamma^+$, 
$\Gamma^{R_2}=\gamma_i^\perp$,
we obtain that
\begin{eqnarray}
\label{Sp-l-F-1}
&& 
\big[ \bar u^{(\up_x)}(k) \gamma^+ \gamma^\perp_j \gamma_5 u^{(\up_x)}(k)\big]
\big[ \bar u^{(\up_x)}(k) u^{(\up_x)}(k)\big]=
\nonumber\\
&&
C\,\big[ \bar u^{(\up_x)} (k) \gamma^+ u^{(\up_x)}(k)\big]
\big[ \bar u^{(\up_x)}(k) \gamma^\perp_i u^{(\up_x)}(k)\big].
\end{eqnarray}
Here, we assume that, for the fixed indices, $i\not= j$, the coefficient $C$ is given by 
\begin{eqnarray}
\label{C-F-T}
C= \frac{1}{16} \text{tr}\big[ \gamma^\perp_i \gamma^+ \gamma^\perp_j \gamma_5 \gamma^-\big]
\text{tr}\big[ \gamma^\perp_i \gamma^\perp_i \big].
\end{eqnarray}
Eqn.~(\ref{Sp-l-F-1}) can be readily inverted in order to get the following 
representation of Eqn.~(\ref{Me-4})
\begin{eqnarray}
\label{Me-5}
&& 
\big[ \bar u^{(\up_x)}(k) \gamma^+ \gamma^\perp_1 \gamma_5 u^{(\up_x)}(k)\big]
\int (d^4 \tilde k) 
\int (d^4 \ell) 
\nonumber\\
&&
\times
\Delta(\ell^2)  \Delta(k^2)
{\cal M}\left( k^2, \ell^2, \tilde k^2, ... \right)\Rightarrow
\nonumber\\
&&
k^+ i \epsilon^{+ - P_\perp s_\perp} \tilde f_1^{(1)} (x; \,k^2_\perp).
\end{eqnarray}
On the {\it r.h.s.} of (\ref{Me-5}), we write down the complex $i$ explicitly in order to stress that
it stems from the trace of four $\gamma$-matrices with $\gamma_5$.
For the existence of Lorentz vector defined as $\epsilon^{+ - P_\perp s_\perp}$, it is necessary to assume that 
the quark spin $s_\perp$ is not a collinear vector to the hadron transverse momentum $P_\perp$. 
Within the CS-frame, the hadron transverse momentum can be naturally presented as 
$\vec{\bf P}_\perp=(P_1^\perp, 0)$. Since the hadron spin vector $S$ can be decomposed on the 
longitudinal and transverse components as 
$S^L + S^\perp = \lambda P^+/m_N + S^\perp$,
we get $P\cdot S = \vec{\bf P}^\perp\,\vec{\bf S}^\perp =0$. Hence, in the CS-frame, it is natural to suppose that 
quark $s^\perp$ and hadron $S^\perp$ are collinear ones. 
This is a kinematical constraint (or evidence) for the nonzero Lorentz combination $\epsilon^{+ - P_\perp s_\perp}$
and, therefore, for the existence of a new function $ \tilde f_1^{(1)} (x; \,k^2_\perp)$. 

We have reached the inference that Eqn.~(\ref{Me-5}) contains the Lorentz structure with the quark polarization vector $s_\perp$. 
Therefore, within the frame of the ``nonstandard way'',  the Lorentz vector defined as 
$\epsilon^{+ - P_\perp s_\perp}$ should be included in the general scheme 
of the Lorentz parametrization applied for the relevant correlators. 

Notice that the standard
\footnote{The taking into account of the final(initial) state interaction in the corresponding in(out)-states may change 
the conclusion of the functional complexity. This subtlety is not discussing in the present study.}
time-reversal transformation and the Hermitian conjugation suggest that the function $\tilde f_1^{(1)} (x; \,k^2_\perp)$ 
is a pure imaginary and T-odd function.
We emphasize that the $k_\perp$-dependent parton functions may possess 
the nontrivial properties under the replacement $k_\perp \to -k_\perp$
owing to that the time-reversal transforms convert the future-pointed WL to the past-pointed WL \cite{Boer:2003cm}.

It is important that the combination $\epsilon^{+ - P_\perp s_\perp}$
is not the only new combination. For example, we may construct an analogous combination defined as 
$\epsilon^{+ - k_\perp s_\perp}$. Actually, the ``nonstandard'' way presented in the paper discovers a new kind of  
different $k_\perp$-dependent functions. The comprehensive analysis 
of a such kind of functions is planned to be done in the forthcoming works. 

So, it explicitly shows that the function $\tilde f_1^{(1)} (x; \,k^2_\perp)$ and its analogues must appear in the 
parametrization of the hadron matrix element, {\it i.e.} 
\begin{eqnarray}
\label{Phi-plus}
&&\Phi^{[\gamma^+]}(k)\equiv
\int (d^4 z) e^{+i(kz)} \langle P,S | \bar\psi(0) \, \gamma^+ 
\, \psi(z) \, \mathbb{S}[\psi,\bar\psi, A] \,
| P,S \rangle \Big|_{k^+=xP^+} 
\nonumber\\
&&
= i \epsilon^{+ - P_\perp s_\perp} \tilde f_1^{(1)} (x; \,k^2_\perp) + 
i \epsilon^{+ - k_\perp s_\perp} f_{(2)}(x; \,k^2_\perp) + ....,
\end{eqnarray} 
where the ellipse denotes the other possible terms of parametrization \cite{AS-2022}.

Comparing the ``standard'' (the Lorentz parametrization with $\mathbb{S}=\mathbb{I}$ in the correlator) and the ``nonstandard'' 
(the Lorentz parametrization with $\mathbb{S}$-matrix in the correlator) ways, we see that 
the quark spinor lines play a crucial role in the construction of possible Lorentz combination.
Indeed, it can be demonstrated by the following reasoning.
Let us return to Eqn.~(\ref{Phi-1-4}), we are able to specify the corresponding correlator as 
\begin{eqnarray}
\label{Demo-1}
&&\Phi^{[\gamma^+]}(k)\Big|_{\mathbb{S}[\psi,\bar\psi, A]=\mathbb{I}} 
\Longrightarrow \langle P,S | \bar\psi(0) \, 
\gamma^+\, \psi(z) | P,S \rangle =
\nonumber\\
&&
\int (d^4 k_1) (d^4 k_2) e^{-ik_2 z } \,L^{[\gamma^+]} (k_1, k_2) \,
\langle P,S | b^+(k_1) \, b^-(k_2) | P,S \rangle,
\end{eqnarray}
where the spinor line function $L^{[\gamma^+]} (k_1, k_2) $ reads
\begin{eqnarray}
\label{L-form}
L^{[\gamma^+]} (k_1, k_2) = \bar u(k_1) \, \gamma^+ \, u(k_2).
\end{eqnarray}
It is clear that if the correlator involves the interactions generated by the corresponding higher orders of $\mathbb{S}$-matrix, 
the spinor line function $L^{[\Gamma]} (k_1, k_2)$ becomes more complicated in comparison with Eqn.~(\ref{L-form}).
That is, we have 
\begin{eqnarray}
\label{Demo-2}
&&\Phi^{[\gamma^+]}(k) 
\Longrightarrow \langle P,S | \bar\psi(0) \, 
\gamma^+\, \psi(z) \, \mathbb{S}^{(n)}[\psi,\bar\psi, A] \, | P,S \rangle =
\nonumber\\
&&
\int d^4 \mu(k_1, ..., k_m)  \,L^{[\Gamma_1 ... \Gamma_N]} (k_1,... k_N) \,
\langle P,S | b^+(k_1) ... b^+(k_i)\, b^-(k_{i+1}) ... b^-(k_N) | P,S \rangle,
\end{eqnarray}
where the spinor line function $L^{[\Gamma_1 ... \Gamma_N]} (k_1,... k_N) $ is determined 
by the $N$ spinor lines.
As a result, we have a possibility to introduce the new Lorentz combination 
for parametrization.

Eqn.~(\ref{Phi-plus}) represents our principal result which reveals the existence of
a new transverse-momentum dependent function $\tilde f_1^{(1)} (x; \,k^2_\perp)$ and its analogue $f_{(2)}(x; \,k^2_\perp)$.
We also observe that Lorentz structure tensor, $\epsilon^{+ - P_\perp s_\perp}$, 
associated with our function (see (\ref{Phi-plus}))
resembles the Sivers structure, $\epsilon^{+ - P_\perp S_\perp}$ in which
the nucleon spin vector $S_\perp$ is replaced by the quark spin vector $s_\perp$.
However, despite this similarity the Sivers function and the introduced function  $\tilde f_1^{(1)} (x; \,k^2_\perp)$
have totally  different physical meaning.

Last but not least, the structure function $\tilde f_1^{(1)} (x; \,k^2_\perp)$ of Eqn.~(\ref{Phi-plus}), 
roughly speaking, coincides
with the function $f_1^{(1)} (x; \,k^2_\perp)$ of Eqn.~(\ref{Me-4})
provided {\it (a)} we decipher the hadron matrix element of 
quark(-gluon) operators at least up to the order of $g^2$; {\it (b)} we focus on the regime of $\ell \ll |\tilde k| \sim |k|$  
at this order; {\it (c)} the occurred four spinors generated by
two spinor lines are aligned along the same 
transverse direction.

\section{The Drell-Yan process and the new functions}
\label{DY}

We are now in a position to discuss 
the possible contribution of new functions to different observables.
The simplest example of application 
is related to the well-known unpolarized Drell-Yan (DY) process, {\it i.e.} 
the lepton-production in nucleon-nucleon collision:
\begin{eqnarray}
\label{DY-proc-st}
&&N(P_1) + N(P_2) \to \gamma^*(q) + X(P_X)
\\
&&\hspace{2.65cm}\to\ell(l_1)+\bar\ell(l_2) + X(P_X),
\nonumber
\end{eqnarray}
with the initial unpolarized nucleons $N$.
The importance of the unpolarized DY differential cross section is due to
the fact that it has been involved in the denominators of any spin asymmetries.

As mentioned above, the naive time-reversal transformation together with the Hermitian conjugation imply that 
the new functions are T-odd functions. Therefore, at the leading order, the hadron tensor which 
describes the unpolarized DY-process takes the following form:
\begin{eqnarray}
\label{h-t-4-1}
{\cal W}_{\mu\nu}^{(0)}= &&\delta^{(2)}(\vec{\bf q}_\perp) 
\int (d x)  (d y) \delta(x P^{+}_1 - q^+)
\delta(yP^{-}_2 -q^-)
\nonumber\\
&&
\times
{\rm tr} \big[ \gamma_\nu \, \gamma^+ \,\gamma_\mu \, \gamma^- \big]\,
\Phi^{[\gamma^-]}(y) \Big\{ \int (d^2 \vec{\bf k}_1^\perp )\bar\Phi^{[\gamma^+]}(x, k^{\perp\,2}_1) \Big\},
\end{eqnarray}
where 
\begin{eqnarray}
\label{DY-funs}
\Phi^{[\gamma^-]}(y) = P_2^-\, f(y), \quad 
\bar\Phi^{[\gamma^+]}(x, k^{\perp\,2}_1) = 
i\epsilon^{+ - k_1^\perp s^\perp} f_{(2)} (x; \,k^{\perp\, 2}_1).
\end{eqnarray}
This expression is derived within the factorization procedure described in detail in \cite{Anikin:2015xka}.
We stress that we adhere the CS-kinematics \cite{Barone:2001sp} in which
the factorization procedure has the most archetypal form \cite{AS-2022}.

Calculating the contraction of Eqn.~(\ref{h-t-4-1}) with the unpolarized lepton tensor ${\cal L}^U_{\mu\nu} $, we derive that
\begin{eqnarray}
\label{xsec-unpl}
&&d\sigma^{unpol.} \sim \int (d^2 \vec{\bf q}_\perp) {\cal L}^{U}_{\mu\nu} 
{\cal W}^{(0)}_{\mu\nu} =  
\nonumber\\
&&
\int (d x)  (d y) \delta(x P^{+}_1 - q^+)
\delta(yP^{-}_2 -q^-)
\nonumber\\
&&
\times
(1+\cos^2\theta) f(y)  \int (d^2 \vec{\bf k}_1^\perp ) \epsilon^{P_2 - k_1^\perp s^\perp} \Im{m} f_{(2)} (x; \,k^{\perp\, 2}_1),
\end{eqnarray}
where 
\begin{eqnarray}
\epsilon^{+ - k_1^\perp s^\perp} = \vec{\bf k}_1^\perp \wedge \vec{\bf s}^{\perp} \sim \sin (\phi_k - \phi_s)
\end{eqnarray}
with $\phi_A$, for $A=(k, s)$, denoting the angles between $\vec{\bf A}_\perp$ and $O\hat x$-axis in the CS-frame.  

Notice that the angle $\phi_s$ cannot explicitly be measured in the experiment. 
However, the implementation of the covariant (invariant) integration of $f_{(2)} (x; \,k^{\perp\, 2}_1)$  gives the 
kinematical constraints on this angle relating the quark spin angle to the corresponding hadron angle. 
Indeed,  let us consider the integration of Eqn.~(\ref{xsec-unpl}) given by
\begin{eqnarray}
\label{App-Inv-Int-5}
{\cal I}^\perp_\alpha=
\int (d^2 \vec{\bf k}_1^\perp) \, f_{(2)}(x, k_1^\perp, s_\perp ; P_1) \, k^\perp_{1\, \alpha}.
\end{eqnarray}
Based on the Lorentz covariance, the integration ${\cal I}^\perp_\alpha$ can be presented as 
\begin{eqnarray}
\label{Cov-Int-2}
{\cal I}^\perp_\alpha= P^\perp_{1\, \alpha} \, \mathbb{A}  +
i\,\varepsilon_{\alpha s_\perp + -} \mathbb{B}
\end{eqnarray}
where $P^\perp_{1\, \alpha}$ and $\varepsilon_{\alpha s_\perp + -}$ form the orthogonal system
and, hence, we have 
\begin{eqnarray}
\label{B-App}
&&
\mathbb{A}=\int (d^2 \vec{\bf k}_1^\perp) \, f_{(2)}(x, k_1^\perp, s^\perp ; P_1) \, 
\frac{(k^\perp_1 \cdot P^\perp_1)}{P^{\perp\,2}_1},
\\
&&
\label{C-App}
\mathbb{B}=\int (d^2 \vec{\bf k}_1^\perp) \, f_{(2)}(x, k_1^\perp, s_\perp ; P_1) \, \frac{-i\,\varepsilon^{k_1^\perp s_\perp + -}}{s^2_\perp}.
\end{eqnarray}
Since $P^\perp_{1\, \alpha}$ and $\varepsilon_{\alpha s_\perp + -}$ are orthogonal each other by construction, 
we have that  
\begin{eqnarray}
\label{App-Req-1}
\varepsilon^{P^\perp_1 s^\perp + -}\sim \sin\varphi_{Ps} =0 \Rightarrow \varphi_P = \varphi_s \pm n\pi.
\end{eqnarray}
In other words, one can see that the orthogonality condition required by the covariant integration leads to 
the (anti)collinearity of $P^\perp_1$ and $s^\perp$. Eqn.~(\ref{App-Req-1}) relates the hadron momentum with the 
quark spin vector and it can be treated as another condition for the existence of this new function.

Thus, the new $k_\perp$-dependent function $f_{(2)} (x; \,k^{\perp\, 2}_1)$ gives the additional and additive contribution to the 
depolarization factor $D_{[1+\cos^2\theta]}$ appeared in the differential cross section of unpolarized DY process.

\section{Conclusions}
\label{Con}

To conclude, in the paper we have introduced the new $k_\perp$-dependent function $\tilde f_1^{(1)} (x; \,k^2_\perp)$
and $f_{(2)} (x; \,k_{\perp}^{2})$ of Eqn.~(\ref{Phi-plus})
which describe the transverse quark motion by the quark alignment along the fixed transverse direction.
The introduced functions can be considered as a ``in-between'' functions 
of the Sivers and Boer-Mulders functions. Indeed, the Lorentz tensors accompanying our functions 
are quite similar to the analogous tensor at the Sivers function, however they describe the polarized quark effects inside
the unpolarized nucleon like the Boer-Mulders function.  

We have shown that, to the second order of strong interactions, 
the new parametrizing function $\tilde f_1^{(1)} (x; \,k^2_\perp)$ can be related to
the function $f_1^{(1)} (x; \,k^2_\perp)$ of (\ref{Me-1}) imposing 
the condition $\ell \ll |\tilde k| \sim |k|$ which corresponds to the regime where  
the appeared two spinor lines are interacting by exchanging of soft gluon. 
Moreover, the occurred four spinors generated by
two spinor lines have the polarizations aligned along the same 
transverse direction. 
In physical terms, the $k_\perp$-dependent function $\tilde f_1^{(1)} (x; \,k^2_\perp)$ which describes 
the regime where
$k_\perp$-dependence (or the transverse motion of quarks inside the hadron) has been 
entirely generated by the 
quark spin alignment.
From the mechanical point of view, it resembles the 
deviation of alike-rotated balls from the straightforward motion.

As a practical application of the new functions, we have illustrated that the function $f_{(2)}(x; k^2_\perp)$
provides the additional contribution to the depolarization factor $D_{[1+\cos^2\theta]}$ 
which is associated with the differential cross section of unpolarized DY process.

\acknowledgments

We thank colleagues from the Theoretical Physics Division of NCBJ (Warsaw)
for useful and stimulating discussions.
The work of L.Sz. is supported by the grant 2019/33/B/ST2/02588  
of the National Science Center in Poland.
This work is also supported by the Ulam Program of NAWA No.
PPN/ULM/2020/1/00019.


\appendix
\renewcommand{\theequation}{\Alph{section}.\arabic{equation}}

\section{The amplitude and the interaction (or Heisenberg) representation}
\label{App-Amp}

In this appendix we remind a role of the interaction representation in
the parametrization of relevant nonperturbative correlators,
despite the subject is the well-known textbook.

For example, we begin with the forward Compton scattering (CS) amplitude which is related to the deep inelastic scattering.
It reads
\begin{eqnarray}
\label{Amp-1}
{\cal A}_{\mu\nu} = \langle P| a^-_\nu(q) \, \mathbb{S}[\bar\psi, \psi, A] \, a^+_\mu(q) | P\rangle,
\end{eqnarray}
and we recall that 
\begin{eqnarray}
\mathbb{S}[\psi,\bar\psi, A]=T\text{exp}\Big\{ 
i \int (d^4 z) \big[ {\cal L}_{QCD}(z) + {\cal L}_{QED}(z)\big] \Big\}.
\nonumber
\end{eqnarray}
The commutation relations of creation (or annihilation) operators with $\mathbb{S}$-matrix are given by 
\begin{eqnarray}
\label{Com-1}
\big[ a^\pm_\mu(q), \, \mathbb{S}[\bar\psi, \psi, A] \big]= \int (d^4 \xi) e^{\pm i q \xi} 
\frac{\delta \mathbb{S}[\bar\psi, \psi, A] }{ \delta A^\mu(\xi)},
\end{eqnarray}
where
\begin{eqnarray}
\label{Der-S}
\frac{\delta \mathbb{S}[\bar\psi, \psi, A] }{ \delta A^\mu(\xi)} = 
T\Big\{  \int (d^4 z) \frac{\delta {\cal L}_{QED}(z)}{\delta A^\mu(\xi)} \, \mathbb{S}[\bar\psi, \psi, A] \Big\}.
\end{eqnarray}
Having used Eqn.~(\ref{Com-1}) and the translation invariance, the CS-amplitude takes the form of 
\begin{eqnarray}
\label{Amp-2}
&&{\cal A}_{\mu\nu} = 
\int (d^4 \xi_1) (d^4 \xi_2) e^{- i q (\xi_1 - \xi_2)} 
\langle P| \frac{\delta^2 \mathbb{S}[\bar\psi, \psi, A] }{ \delta A^\mu(\xi_1) \delta A^\nu(\xi_2)} | P\rangle
\nonumber\\
&&
\Rightarrow \int (d^4 z) e^{- i q z} \langle P| T \Big( J_\nu(0) \, J_\mu(z) \mathbb{S}[\bar\psi, \psi, A]  \Big)| P\rangle.
\end{eqnarray}
From Eqn.~(\ref{Amp-2}), using Wick's theorem we can readily derive the ``hand-bag'' diagram contribution 
which has a form of (in the momentum representation before the factorization procedure applied)
 \begin{eqnarray}
\label{Amp-3}
&&{\cal A}_{\mu\nu} = 
\int (d^4 k) \,\text{tr} \Big[ E_{\mu\nu}(k) \Phi(k) \Big],
\end{eqnarray}
where 
\begin{eqnarray}
\label{E}
&&E_{\mu\nu}(k) = \gamma_\mu S(k+q) \gamma_\nu + \gamma_\nu S(k-q) \gamma_\mu,
\\
&&
\label{phifun}
\Phi(k)= \int (d^4 z) \, e^{ikz} 
\langle P| T \bar\psi(0) \psi(z) \mathbb{S}[\bar\psi, \psi, A] | P \rangle_c,
\end{eqnarray}
where the subscript $c$ denotes the connected diagram contributions only.
Notice that after the factorization procedure applied to the CS-amplitude, $\mathbb{S}$-matrix has to be included 
in the corresponding correlator forming the soft part of amplitude. 

It is worth to notice that in Eqn.~(\ref{phifun}) the nonperturbative correlator has been written in the interaction representation.
In the literature, the wide-used Heisenberg representation of this correlator, {\it i.e.}
\begin{eqnarray}
\label{phifun-H}
\Phi(k)= \int (d^4 z) \, e^{ikz} 
\langle P| \bar\psi(0) \psi(z) | P \rangle^H,
\end{eqnarray} 
gives a very compact form but it may misleads, however, in many cases.
Indeed, the neglecting $H$ in Eqn.~(\ref{phifun-H}) results in the wrong impression about the absence of interaction 
in the correlator.  



\begin{thebibliography}{99}
\vspace{1\baselineskip}

\bibitem{Boglione:1999pz}
M.~Boglione and P.~J.~Mulders,
Phys. Rev. D \textbf{60}, 054007 (1999)

\bibitem{Bacchetta:2004jz}
A.~Bacchetta, U.~D'Alesio, M.~Diehl and C.~A.~Miller,
Phys. Rev. D \textbf{70}, 117504 (2004)

\bibitem{Goeke:2005hb}
K.~Goeke, A.~Metz and M.~Schlegel,
Phys. Lett. B \textbf{618}, 90-96 (2005)

\bibitem{Collins:2005rq}
J.~C.~Collins,A.~V.~Efremov, K.~Goeke, M.~Grosse Perdekamp, S.~Menzel, B.~Meredith, A.~Metz and P.~Schweitzer,
Phys. Rev. D \textbf{73}, 094023 (2006)

\bibitem{Anselmino:2008sga}
M.~Anselmino, M.~Boglione, U.~D'Alesio, A.~Kotzinian, S.~Melis, F.~Murgia, A.~Prokudin and C.~Turk,
Eur. Phys. J. A \textbf{39}, 89-100 (2009)

\bibitem{Bastami:2018xqd}
S.~Bastami, H.~Avakian, A.~V.~Efremov, A.~Kotzinian, B.~U.~Musch, 
B.~Parsamyan, A.~Prokudin, M.~Schlegel, G.~Schnell and P.~Schweitzer, \textit{et al.}
JHEP \textbf{06}, 007 (2019)

\bibitem{Boer:1997nt}
D.~Boer and P.~J.~Mulders,
Phys. Rev. D \textbf{57}, 5780-5786 (1998)

\bibitem{Itzykson:1980rh}
C.~Itzykson and J.~B.~Zuber,
``Quantum Field Theory,'' 
International Series In Pure and Applied Physics.

\bibitem{Barone:2001sp}
V.~Barone, A.~Drago and P.~G.~Ratcliffe,
Phys. Rept. \textbf{359}, 1-168 (2002)

\bibitem{Boer:2003cm}
  D.~Boer, P.~J.~Mulders and F.~Pijlman,
  Nucl.\ Phys.\ B {\bf 667}, 201 (2003)

\bibitem{Anikin:2017azc}
I.~V.~Anikin, L.~Szymanowski, O.~V.~Teryaev and N.~Volchanskiy,
Phys. Rev. D \textbf{95}, no.11, 111501 (2017)

\bibitem{Anikin:2015xka}
I.~V.~Anikin and O.~V.~Teryaev,
Eur. Phys. J. C \textbf{75}, no.5, 184 (2015)

\bibitem{AS-2022}
I.~V.~Anikin, L.~Szymanowski,
in preparation

\end{thebibliography}
\end{document}